\documentclass[12pt,preprint]{aastex}
\usepackage{graphicx}
\usepackage{amssymb,amsmath}

\pdfoutput=1

\newcommand\beq{\begin{equation}}
\newcommand\eeq{\end{equation}}
\newcommand\lsim{\mathrel{\rlap{\lower4pt\hbox{\hskip1pt$\sim$}}
        \raise1pt\hbox{$<$}}}
\newcommand\gsim{\mathrel{\rlap{\lower4pt\hbox{\hskip1pt$\sim$}}
        \raise1pt\hbox{$>$}}}

\begin{document}
\title{Water-Trapped Worlds}

\author{Kristen
  Menou\altaffilmark{1}}
 
\altaffiltext{1}{Department of Astronomy, Columbia University, 550
  West 120th Street, New York, NY 10027}

\begin{abstract}
Although tidally-locked habitable planets orbiting nearby M-dwarf
stars are among the best astronomical targets to search for extrasolar
life, they may also be deficient in volatiles and water. Climate
models for this class of planets show atmospheric transport of water
from the dayside to the nightside, where it is precipitated as snow
and trapped as ice. Since ice only slowly flows back to the dayside
upon accumulation, the resulting hydrological cycle can trap a large
amount of water in the form of nightside ice.  Using ice sheet
dynamical and thermodynamical constraints, I illustrate how planets
with less than about a quarter the Earth's oceans could trap most of
their surface water on the nightside. This would leave their dayside,
where habitable conditions are met, potentially dry. The amount and
distribution of residual liquid water on the dayside depend on a
variety of geophysical factors, including the efficiency of rock
weathering at regulating atmospheric CO$_2$ as dayside ocean basins
dry-up. Water-trapped worlds with dry daysides may offer similar
advantages as land planets for habitability, by contrast with worlds
where more abundant water freely flows around the globe.
\end{abstract}

\section{Introduction}

Tidally-locked terrestrial planets orbiting in the habitable zone of
nearby M-dwarf stars are among the best astronomical targets for
discovery and atmospheric characterization (e.g., Charbonneau \&
Deming 2007; Montgomery \& Laughlin 2009; Seager \& Deming 2010;
Rojas-Ayala et al. 2013). Recent estimates based on Kepler results are
encouraging in suggesting a high frequency of such planets around
M-dwarfs (Dressing \& Charbonneau 2013; Fressin et al. 2013; Kopparapu
2013; Morton \& Swift 2013).

The climate on this class of planets is peculiar and interesting
because of the permanent day-night insolation pattern many are
expected to experience by the end of their tidal evolution, once they
are captured in 1:1 spin-orbit resonance (e.g. Rodriguez et al. 2012;
see Correia \& Laskar 2011 for a review). Provided that large enough
surface pressures are present, nightside atmospheric collapse can be
avoided (Joshi et al. 1997). Various other studies have illustrated
how habitable planets around M-dwarfs will typically find themselves
in a climate regime with hot daysides, cold frozen nightsides and an
atmospheric circulation connecting the two hemispheres (Joshi 2003;
Merlis \& Schneider 2010; Heng et al. 2011; Kite et al. 2011;
Pierrehumbert 2011; Wordsworth et al. 2011; Edson et al. 2011,2012;
Leconte et al. 2013).

Two aspects of the climate problem on these planets which have
received less attention, however, are the possibilities that they are
deficient in volatiles and that a significant fraction of their
surface water inventory could be trapped as ice on their nightside.
Indeed, Lissauer (2007) and Raymond et al. (2007) have argued on the
basis of standard terrestrial planet formation scenarios that volatile
deficiency may be a key attribute of these planets (see also Ogira \&
Ida 2009 for an alternate view). Even planets formed with abundant
water could lose much of their surface inventory via atmospheric
escape during an intense tidal heating phase preceding their
spin-orbit resonant capture (Barnes et al. 2013). Heath et al. (1999)
and Joshi (2003) have considered the nightside ice trap issue but they
mostly emphasized how the ice layer would experience melting at its
base on worlds with abundant water, thus making the ice trap
inefficient. Recently, Leconte et al. (2013) have also discussed the
nightside ice trap but mostly in the specific context of the climate
bi-stability of planets found near the inner edge of their habitable
zone.

Here, I interpret volatile deficiency as likely implying a small
surface water inventory for these planets and I argue that the ice
trap on the nightside of water-deficient planets around M-dwarfs could
in fact be efficient and have a profound effect on their climate. In
\S\ref{sec:climate}, I present new simulations of the climate on
habitable planets around M-dwarfs which illustrate the rapid
development of a nightside ice trap for surface water. In
\S\ref{sec:ice}, I develop the argument about the effective nature of
this ice trap on sufficiently water-deficient planets, before
discussing some implications for their climate and concluding in
\S\ref{sec:conclusion}.

\section{Climate Simulations and Ice Trapping} \label{sec:climate}

I use PlanetSimulator, a flexible Earth-System simulator of
intermediate complexity developed at the University of
Hamburg,\footnote{www.mi.uni-hamburg.de/plasim} to study the climate
of tidally-locked habitable planets around M-dwarfs. PlanetSimulator
relies on a pseudo-spectral atmospheric dynamical solver coupled to an
accurate radiative transfer scheme. It includes detailed formulations
to simulate the hydrological cycle on Earth-like planets, such as
water evaporation and precipitation, diagnostic cloud formation, a 50m
slab model for the ocean and a thermodynamic sea ice model (see
Fraedrich et al. 2005 for details). Although these model elements have
been tuned to simulate the Earth's climate, they are expected to
remain reasonably accurate for climates deviating only modestly from
Earth's conditions, as is the case here. I use PlanetSimulator in its
T42L10 formulation (128 longitude nodes, 64 latitude nodes and 10
vertical levels).

I modified PlanetSimulator to enforce a permanent day-night insolation
pattern and I set the planetary obliquity and orbital eccentricity to
zero in all models, for consistency with the tidally-locked hypothesis
investigated here. I adopt an Earth-like insolation of
$1365$~W~m$^{-2}$ in all models and a planet rotation period $P_{\rm
  rot} =240$~h by default, which corresponds to a tidally-locked
planet around a $\sim 0.3$~$M_\sun$ M-dwarf (Edson et al. 2012).  I
also removed orography (relief over continents) from all models to
make the results more general and after tests showed that it does not
have a major impact on the main results. I use the current
distribution of Earth continents in models with continents.  All other
model parameters take their standard Earth values, unless otherwise
stated.\footnote{In particular, I have not accounted for the reduced
  ice and snow albedos expected on planets orbiting M-dwarfs (Joshi \&
  Haberle 2012). This is expected to have only a minor quantitative
  impact on the main results of interest here because only a small
  fraction of the icy surface is actually exposed to insolation (see
  also Leconte et al. 2013).}

Table~\ref{tab:one} shows the list of models I have explored in this
work. All models are run for 40 Earth months, which is sufficient for
the surface temperature field to reach a steady-state under permanent
insolation, starting from Earth-like initial conditions. Earth1,
Earth2 and Earth3 are models with standard Earth parameters, the
default rotation period $P_{\rm rot} =240$~h and a permanent
insolation pattern centered on longitude $0$~deg (Greenwhich),
$+90$~deg (Bangladesh) and $+180$~deg (central pacific ocean),
respectively.  Earth.lowP is similar to Earth1 except that the total
atmospheric surface pressure is reduced to
$0.3$~bar.\footnote{Volatile deficiency could in principle also result
  in a reduce N$_2$ atmospheric content, which is captured by the
  lower surface pressure in this model. Note that improved modeling of
  low surface pressure scenarios would entail verifying the accuracy
  of the PlanetSimulator radiative scheme under these conditions,
  which was not done here.} With a constant CO$_2$ fraction of
$360$~ppm, this also corresponds to a CO$_2$ atmospheric content
reduced by about a third. Earth.lowCO2 and Earth.highCO2 are similar
to Earth1 except for CO$_2$ fraction of $36$~ppm and $3600$~ppm,
respectively.\footnote{Lucarini et al. (2013) suggest that
  PlanetSimulator's radiative transfer scheme should remain reasonably
  accurate over this range of CO$_2$ atmospheric content values.}
Earth.fast and Earth.slow are similar to Earth1 except for their
rotation periods $P_{\rm rot} =120$~h and $480$~h,
respectively. SuperEarth is similar to Earth1 except that it uses
larger values of the surface gravity and the planetary radius,
corresponding to a super-Earth with a mass $\sim 10$~$M_\earth$ (see
Table~\ref{tab:one}; Valencia et al. 2006). Finally, Aquaplanet is
similar to the Earth1-3 models except that all the continents have
been replaced by a global ocean.

I find that the general circulation regime in all these models is
qualitatively consistent with what has been reported in the literature
before for other tidally-locked habitable planet models (e.g., Joshi
et al. 1997; Merlis \& Schneider 2010; Heng et al. 2011; Edson et
al. 2012).  The majority of models listed in Table~\ref{tab:one}
exhibit a zonally-averaged zonal wind profile comparable to that of
Earth, with two eastward jets streams at altitude and negative wind
speeds at the equatorial ground. In three models, however (Earth.lowP,
Earth.slow and SuperEarth), the general circulation is in a
superrotating\footnote{Superrotation refers to a regime with positive
  equatorial winds, which imply ``eddy'' momentum transport towards
  the equator beyond what any type of axisymmetric meridional
  circulation can achieve (e.g., Showman et al. 2011).} regime with
positive zonal winds at the equatorial ground and a rather broad
eastward jet at altitude. Maximum zonally-averaged wind speeds range
from $\sim 17$~m~s$^{-1}$ (SuperEarth model) to $\sim 60$~m~s$^{-1}$
(Earth.highCO2). Interestingly, these results are not entirely
consistent with the study of Edson et al. (2012), who reported faster
wind speeds in their superrotating models and dynamical transitions to
the superrotating regime at $P_{\rm rot} \simeq 100$--$101$~h for land
planets and $P_{\rm rot} \simeq 72$--$96$~h for aquaplanets (i.e., at
faster rotation rates than found here). It would be interesting to
understand the origin of this discrepancy better. I simply note for
now that it does not appear to have a major impact on the main results
of interest here for nightside ice since all models in
Table~\ref{tab:one} exhibit comparable properties in that respect.

Figure~\ref{fig:one} shows representative end results from the
Aquaplanet model. The surface temperature field shows that liquid
water conditions are confined to the dayside of the planet (see also
Figure~\ref{fig:four}), making this model climate comparable to the
eyeball climate discussed by Pierrehumbert (2011). On the nightside,
the coldest surface conditions are offset from the poles and the
equator is the hottest, as the region where the near-surface wind is
the most efficient at carrying heat from the dayside and coupling it
with the nightside surface.

Precipitable water in the atmosphere is also largely confined to the
dayside, where evaporation strongly peaks at the substellar point. The
circulation is able to advect, preferentially eastward, a modest
amount of moisture to the nightside, where it is precipitated as
snow. Most of the snow precipitation occurs in an annulus on the
dayside, but the weak snow precipitation rates on the nightside are
nevertheless interesting in terms of the ice accumulation rates they
imply (see below). As noted before by Merlis \& Schneider (2010) in
comparable simulations, strong net precipitation occurs in the
vicinity of the substellar point, with net evaporation over most of
the dayside.  All of the properties illustrated in
Figure~\ref{fig:one} are qualitatively shared by the other models
listed on Table~\ref{tab:one}, which suggest that they are rather
general for tidally-locked habitable planets around M-dwarf stars.

For each of the models, Table~\ref{tab:one} also lists the snow
precipitation rate averaged over the planetary nightside. Values range
from $0.3$ to $0.01$~mm~day$^{-1}$, in broad agreement with the
$0.11$~mm~day$^{-1}$ nightside rate reported by Joshi
(2003).\footnote{Note that Joshi's simulations are not fully
  consistent for M-dwarf habitable planets in that they assume an
  Earth-like planetary rotation rate ($P_{\rm rot}=24$~h).}  This
range largely reflects the efficiency of the circulation at advecting
moisture to the nightside, which is noticeably weaker in the
Earth.slow model (see also Merlis \& Schneider 2010).  Nevertheless,
with such precipitation rates, a layer of $1$~km of ice would take
between $3 \times 10^4$ and $10^6$~yr, and typically $\sim 10^5$~yr,
to accumulate on the nightside (accounting for the factor $\sim 3$
difference between the density of snow and ice). These timescales are
quite short and they suggest a rapid change in the hydrological cycle
of planets with no more than a few km of ocean (such as Earth),
following their capture into 1:1 spin-orbit resonance. It is thus
important to understand the nature of this ice layer in detail if one
is to address the climate of habitable planets around M-dwarfs.

\section{Nightside Ice Sheet} \label{sec:ice}

\subsection{General Considerations}

Figure~\ref{fig:two} illustrates schematically three possible
configurations for the nightside ice layer on a tidally-locked
planet. For the sake of argument, a uniformly flat ocean basin has
been assumed and continents have been ignored, though the consequences
of lifting these assumptions are reconsidered in
\S\ref{sec:conclusion}.

As discussed quantitatively in \S\ref{sec:melting}, a conducting ice
layer carrying a given geothermal heat flux has a limited thickness
because melting eventually occurs at its base.\footnote{At least for
  ice pressures $\ll 100$s~MPa, which is the relevant regime for the
  water-deficient planets of interest here.} This limiting thickness
is smaller for warmer surface temperatures at the top of the ice
layer. Tidally-locked planets with relatively mild nightside
temperatures and/or a large surface water inventory will thus have
their nightside ice layer float over a sub-glacial ocean connecting to
the dayside ocean. This scenario, which is the one emphasized by
Health et al. (1999; see also Joshi 2003), is depicted in the first
``ice shelf configuration'' shown in Fig.~\ref{fig:two}. In this
configuration, rain and snow precipitate on the nightside, as a result
of moisture advection from the dayside (gray arrows). Snowfall does
not add to the ice layer thickness, however, because the thickness is
limited by basal melting. The steady-state hydrological cycle is
closed with a return flow of water to the dayside via the sub-glacial
ocean (yellow arrows). The ice+ocean layer thickness must be
essentially uniform over the entire globe to satisfy lateral pressure
equilibrium at the base, or rapid water flows would ensue to enforce
this equilibrium.\footnote{Note that the thick ice shelf
  representation in Fig.~\ref{fig:two} is highly idealized since ice
  shelves on Earth are known to spread out and thin out efficiently,
  and they can experience catastrophic break-ups (e.g., Tziperman et
  al. 2012, Scambos et al. 2009). }

As depicted by the second``ice sheet configuration'' in
Fig.~\ref{fig:two}, on a planet with less surface water and/or colder
nightside temperatures, it becomes possible for part of the ice layer
not to exceed the critical thickness for melting. This region of the
ice layer is then grounded, as shown in between the two radial dotted
lines in Fig.~\ref{fig:two}. Rain and snow precipitation act the same
as in the previous configuration, with the important exception of the
grounded ice sheet. Ice is able to accumulate in this ice sheet and
sustain a sizable pressure gradient at its base (indicated by a bulged
ice layer in Fig.~\ref{fig:two}) because of the considerable ice
viscosity. Ice flows down the basal pressure gradient in the ice sheet
and, in steady state, the hydrological cycle must satisfy a new mass
balance: over the grounded ice sheet, ice accumulation from snowfall
must balance the flow of ice exiting the ice sheet at its boundary
with the sub-glacial ocean. Since ice accumulation is made possible by
the enormous difference of viscosity between the grounded ice flow (in
the ice sheet) and the water flow (in the ocean) covering the rest of
the globe, an ice sheet model describing the viscous flow of ice is
necessary to evaluate the amount of ice present in the ice sheet.

The third ``water-trapped configuration,'' which is of particular
interest to the present study, is essentially an extreme version of
the ``ice shelf configuration.'' On a planet with even less surface
water and/or even colder nightside temperatures, one can in principle
have the great majority of the surface water trapped in the slowly
flowing ice sheet, with little water present in liquid form on the
dayside. A finite liquid dayside inventory remains a necessity to
close the hydrological cycle. Indeed, it is the evaporation, advection
and subsequent precipitation of this dayside water on the nightside
that must balance the flux of ice flowing back to the dayside, but the
size of this dayside liquid reservoir could in principle be very
small, as depicted in Fig.~\ref{fig:two}. To evaluate how much water
can potentially be trapped on the nightside in this configuration, I
now turn to quantitative ice sheet models.

\subsection{Isothermal Ice Sheet Models}

Thermo-mechanical models of ice sheets exist (e.g. Bueler et al. 2007;
Fowler 2011) but I find it convenient for the present purpose to
consider separately mechanical and thermodynamical constraints to gain
insight into the behavior of nightside ice on a tidally-locked
planet. The use of a simple, isothermal ice sheet model can be
justified by noting that a comparable diffusion equation can be
derived for the non-isothermal case (e.g., Fowler 2011) and that ice
sheets generally exhibit strong vertical shear preferentially near
their bottom boundary, i.e. over a limited range of temperatures. The
evolutionary equation satisfied by a thin isothermal ice sheet with a
flat base and a no slip boundary condition at its bottom is (Fowler
2011)

\begin{equation}
\frac{\partial h}{\partial t}=  {\bf \nabla} \cdot \left[ 2 (\rho g)^3 A_0
\frac{h^{n+2} |{\bf \nabla} h|^{n-1}}{n+2} {\bf \nabla} h \right] + a, \label{eq:ice1}
\end{equation}

where $h$ is the ice thickness, $\rho=917$~kg~m$^{-3}$ is the mean
density of ice, $g$ is the surface gravity,
$A_0=10^{-16}$~Pa~yr$^{-1}$ is a typical value of the flow rate
constant (related to the ice viscosity near the bottom of the ice
sheet), $n=3$ is the typical value used in Glen's ice flow law to
model the non-Newtonian nature of the ice fluid and $a$ is the ice
accumulation rate (assumed constant to model uniform snowfall over the
ice sheet).  This equation is a non-linear diffusion equation for the
ice thickness, $h$, satisfying the conservation of mass and momentum
in the shallow ice approximation ($h$ much less than the horizontal
length-scale of the ice sheet). The quantity in bracket is the ice
flux, with an effective diffusion coefficient that depends strongly on
the ice thickness and its horizontal gradient.

Equation~(\ref{eq:ice1}) admits analytical solutions (e.g., Bueler
2003) and I shall use such a simple solution here. I solve the
adimensional, steady-state axisymmetric version of Eq~(\ref{eq:ice1})
for the standard $n=3$ value in Glen's law,

\begin{equation}
\frac{1}{r} \frac{d} {dr} \left[r s^{5}
  \left(\frac{ds}{dr}\right)^3 \right] + 1
=0, \label{eq:ice2}
\end{equation} 

where $r=R/L$ is the adimensional cylindrical radius from the ice
shelf center (located in $R=0$), $r=1$ marks the outer edge of the
axisymmetric ice sheet (defined by $R=L$ dimensionally), $s=h/Z$ is
the adimensional ice thickness, and

\begin{equation}
Z= \left( \frac{5 L^4 a}{2 (\rho g)^3 A_0} \right)^{1/8} \label{eq:iceZ}
\end{equation} 

is the ice thickness scale factor.

I integrate this equation a first time with respect to $r$ and use the
central boundary condition $ds / dr = 0$ in $r=0$ (the ice
``ridge''). I integrate a second time with respect to $r$ and use a
finite, arbitrary thickness $h_{\rm out}$ for the boundary condition
in $r=1$. This boundary condition guarantees a finite ice flux at the
point where the ice ceases to be grounded and reaches flotation over
water (indicated by the radial dotted lines in the middle panel of
Fig.~\ref{fig:two}), at which point the ice sheet equation ceases to
be valid (fowler 2011).  The resulting dimensional solution for the
ice sheet thickness is

\begin{equation}
h(R)= Z \left[ 4 \left[ \left( \frac{1}{2} \right)^{4/3} - \left(
    \frac{R}{2 L} \right)^{4/3} \right] + \left( \frac{h_{\rm out}}{Z}
    \right)^{8/3}  \right]^{3/8}. \label{eq:iceresult}
\end{equation}

This solution generalizes the singular case $h_{\rm out}=0$ already
known in the literature (e.g., Bueller 2003). The case $h_{\rm out}>0$
is more satisfactory for the present application in that it permits a
finite ice flux at the outer edge (where flotation is reached) to
balance the ice accumulation rate integrated over the entire ice
sheet.

Figure~\ref{fig:three} shows a few ice thickness profiles derived from
Eq.~(\ref{eq:iceresult}).  The solid line is the solution for an ice
sheet of $L=10000$~km in radius, a uniform ice accumulation
rate\footnote{Note that ice accumulates at a rate approximately $3$
  times slower than snow precipitates, in proportion to the ratio of
  ice to snow density.}  $a=0.1$~mm day$^{-1}$ and a boundary
thickness $h_{\rm out}=1$~km.  The dash-dotted line, which is nearly
indistinguishable from the solid line, corresponds to a similar
solution except for $h_{\rm out}=500$~m, while the dotted line
corresponds to a reduced accumulation rate $a=0.01$ mm day$^{-1}$. The
dashed line is the solution for an ice sheet of $L=5000$~km in radius
(with accumulation rate $a=0.1$~mm day$^{-1}$ and $h_{\rm
  out}=500$~m). The corresponding values of the ice thickness scale
factor in these four solutions are $Z=7737$, $7737$, $5802$ and
$5471$~m, respectively.

An exploration of the parameter space of these simple ice sheet
solutions shows that the value of $h_{\rm out}$ has only a small
impact on the global ice profile, as long as $h_{\rm out} \ll Z$. The
scale factor $Z$ itself is a good indicator of the overall ice
thickness across the profile (and thus the total mass $\sim Z L^2$ in
the ice sheet), as illustrated in Fig.~\ref{fig:three}. Stronger
(weaker) ice accumulation rates lead to thicker/steeper
(thinner/shallower) ice profiles, as expected from the ice sheet
adopting a steady-state thickness profile such that the local ice flux
at any radius carries an amount of ice equal to the integrated
accumulation rate within that radius (Eq.~(\ref{eq:ice2})). It is this
mass balance that sets the typical ice thickness in these models.

The climate models discussed in \S\ref{sec:climate} do suggest the
possibility of an ice sheet extending globally over the entire
nightside, which corresponds to an ice sheet radius $L \simeq
10000$~km on an Earth-size planet.\footnote{All else being equal, a
  super-Earth has ice sheets $g^{-3/8}$ times thinner from stronger
  gravity, although a global ice sheet could also be thicker from the
  extra $L^{1/2}$ horizontal scale factor in Eq.~(\ref{eq:iceZ}).}
For ice accumulation rates representatives of the average nightside
precipitation in these same climate models ($\sim 0.01$--$0.3$~mm/day
of snow, see Table~\ref{tab:one}), a typical ice thickness is thus $Z
\sim 5.5$--$7.7$~km for an ice sheet of radius $L=10000$~km (see also
Eq.~(\ref{eq:iceZ})).  Such ice depths would imply a considerable
amount of ice being trapped on the nightside. As we shall now see,
such a massive ice trap may be limited by additional thermodynamical
constraints.

\subsection{Thermodynamic Melting Constraint} \label{sec:melting}

Ice thickness is also limited by the melting condition for ice at high
pressure, at the base of the ice sheet. In its simplest form, this
process is modeled as a vertical, steady-state conduction problem for
the ice layer transporting a specified geothermal flux to the
surface. Here, I adopt for concreteness the formalism of Abbot \&
Switzer (2011), which accounts for the important variation with
temperature of the ice thermal conductivity. The resulting exponential
temperature profile for the steady-state conductive ice layer yields a
maximum thickness before melting,

\begin{equation}
h_{\rm cond}=\frac{A}{F} \ln \left( \frac{T_{\rm melt}}{T_{\rm surf}} \right),  \label{eq:icecond}
\end{equation}

where the water ice conductivity is given by $k(T)=A T^{-1}$,
$A=651$~W~m$^{-1}$, $T$ is the ice temperature in K, $F$ is the
thermal flux being conducted ($\simeq 0.09$~W~m$^{-2}$ for Earth's
geothermal flux),\footnote{The geothermal flux may scale up with
  planet mass to the power one half, so that ice layers would be
  comparatively thinner on super-Earths (Abbot \& Switzer 2011).}
$T_{\rm melt} \simeq 260$~K is a representative high-pressure water
melting temperature and $T_{\rm surf}$ is the temperature boundary
condition at the top of the ice layer. $T_{\rm surf}$ is determined by
surface-atmosphere exchanges which are modeled explicitly in the
climate simulations described in \S\ref{sec:climate}.

It is therefore straightforward to translate a surface temperature map
into a map of maximum ice thickness before melting occurs, according
to Eq.~(\ref{eq:icecond}). An example is shown in Fig.~\ref{fig:four}
for the Aquaplanet model, assuming Earth's geothermal flux for
$F$. This map shows that the melting constraint on the ice thickness
is stringent. Most of the nightside ice has a maximum depth $\sim
1$--$1.5$~km before melting occurs, with a colder region of a few
thousand km in extent allowing for thicker ice (up to $4.3$~km
depth). A first, conservative limit on the amount of ice that can be
trapped on the nightside of a tidally-locked planet can thus be
obtained by considering that this thermodynamic melting constraint
provides an effective limit to ice accumulation. Since most of the
nightside exhibits surface temperatures which may not permit ice
thickness in excess of $1.5$--$2$~km without melting, the considerably
larger values of ice thickness ($5.5$-$7.7$~km) that would otherwise
accumulate according to steady-state ice sheet models may not be
realized in practice, because basal melting would prevent such large
ice buildup.  In this interpretation, the actual ice thickness may be
close to marginal basal melting since steady-state mass balance, as
described by simple isothermal ice sheet models, would systematically
have more ice accumulate, until basal melting limits this
accumulation.

A weaker, alternative limit on the amount of ice trapped on the
nightside of a tidally-locked planet can be obtained by considering
the possibly important role of basal water flows on the energetics of
the ice sheet. When basal melting occurs, it is natural for basal
water to advect heat away from the regions where the ice is thickest,
i.e. down the basal pressure gradient. Such a basal water flow,
carrying heat away from melting regions, can reduce the heat flux
$F_{\rm eff}$ effectively conducted through the overlying ice sheet
(e.g., Cuffey \& Paterson 2010), allowing thicker ice than was
suggested by our use of the simple one-dimensional criterion in
Eq.~(\ref{eq:icecond}) (by virtue of $F_{\rm eff}$ being less than
$F$). Although it is difficult to make quantitative predictions for
this effect without a basal water flow model, a strong hypothesis
sometimes adopted in multi-dimensional ice sheet modeling is to assume
that basal water efficiently escapes from under the ice sheet,
carrying away any excess heat needed for melting (e.g., Hulbe \&
MacAyeal 1999). In such a limit, the ice sheet thickness is no longer
strongly limited by melting but it would instead be determined by the
ice sheet mass balance through accumulation (as modeled by
Eq~(\ref{eq:ice1}) for instance).\footnote{In this interpretation, one
  also implicitly assumes that the thick, wet-based ice does not
  experience sliding because it is trapped by thinner, frozen-base ice
  at the periphery of the ice sheet, and that any basal water escapes
  from under the ice sheet through sufficiently localized sub-glacial
  streams. Regular emptying of this ice reservoir through rapid
  sliding events could in principle occur, by analogy with 'Heinrich'
  events for the Laurentide ice sheet during Earth's last glacial
  period (e.g., MacAyeal 1993).}

Although idealized, these two separate limits for ice thickness, which
we refer to as the strong and weak melting constraints, are useful in
bracketing the possible behaviors of ice on the nightside of a
tidally-locked planet, when basal melting occurs. After integration,
the ice distribution following the strong melting constraint, shown in
Fig.~\ref{fig:four}, amounts to an equivalent depth of $560$~m spread
over the entire globe. Similar averages\footnote{For simplicity, I do
  not differentiate between land and ocean when making such
  averages. The role of continents is reconsidered in
  \S\ref{sec:conclusion}.} for the other climates models listed in
Table~\ref{tab:one} range from $320$ to $770$~m.  This range of values
largely reflects how cold the nightside is in each model, with the
lowest efficiency of heat advection to the nightside found in the
model Earth.low (with a low atmospheric mass).  By contrast, the ice
distributions shown as solid and dotted lines in Fig.~\ref{fig:three},
which could in principle be realized according to the weak melting
constraint, amount to equivalent depths (spread over the entire globe)
of $2400$ and $1806$~m for ice sheets covering the entire nightside on
an Earth-sized planet.

The amount of surface water on Earth (essentially in the oceans)
corresponds to an average depth of $\sim 2700$~m once spread over the
globe (e.g., Charette \& Smith 2010). Adopting the strong melting
constraint, one concludes that the climate models listed in
Table~\ref{tab:one} could trap from a seventh up to almost a third of
the Earth's surface water on their nightsides, with the majority of
models trapping about a quarter to a fifth. Following the alternative
weak melting constraint, when efficient sub-glacial water-flows are
present, one concludes that about $65$-$90$\% of the Earth's surface
water could be trapped on the nightside of an Earth-sized
planet.\footnote{Such large amounts of nightside ice are likely
  overestimated because the corresponding ice sheet thickness, shown
  in Fig.~\ref{fig:three}, approaches the tropospheric height of the
  atmosphere, which would reduce snow precipitation on the
  nightside. To evaluate the magnitude of this effect, which will act
  to reduce the ice thickness, one will have to consistently include
  the orographic blueprint of a thick ice sheet in specially-designed
  climate models.}  In both cases, this is considerably more that the
amount of water in ice form on Earth and large enough that
water-deficient planets may be at risk of trapping most of their
surface water on their nightside.

It is difficult, however, to make robust predictions for the total
amount of water present on terrestrial planets on the basis of planet
formation models.  Simulations of in situ formation by Raymond et
al. (2007) suggest that habitable zone planets around M-dwarfs could
form with as little as two orders of magnitude less bulk water than
Earth, which would likely result in significantly reduced amounts of
surface water as well. On the other hand, Ogira \& Ida (2009) argue
that water could be abundantly present on terrestrial planets in the
habitable zone of M-dwarfs if planetary migration efficiently brings
in bodies from beyond the ice line. Depending on details of the
protoplanetary disk structure and how it influences the planetary
migration process, both scenarios could be at work in nature.  M-dwarf
planets could also lose much of their initial surface water inventory
via atmospheric escape during the intense tidal heating phase that may
precede their spin-orbit resonant capture (Barnes et al. 2013).  I
will adopt the simple view here that at least some terrestrial planets
in the habitable zone of M-dwarfs could be quite deficient in surface
water.

Therefore, among the population of habitable Earth-like planets around
M-dwarfs, those who possess only a fraction of the Earth's surface
water inventory could trap the great majority of their water in the
form of nightside ice.  Although the dayside of such water-trapped
worlds may end up being quite dry, it is worth emphasizing again that
a closed hydrological cycle requires liquid water to remain present on
the dayside to compensate for the return flow of ice from the
nightside (see Fig~\ref{fig:two} and discussion below). Evaluating
exactly how much liquid water is present on the dayside of a
water-trapped world is challenging\footnote{For example, Leconte et
  al. (2013) have presented an idealized model for a water reservoir
  present at the edge of the icy region, without accounting for
  constraints on the nightside ice flow.} and, as I shall briefly
describe below, it is closely related to other long-term climate
issues for such planets.

\section{Discussion and Conclusion} \label{sec:conclusion}

Using a combination of climate and ice sheet models, I made the case
for tidally-locked habitable planets around M-dwarfs being able to
trap most of their surface water in the form of nightside ice,
provided they are sufficiently volatile-deficient. Quantitatively,
planets subjected to an Earth-equivalent insolation and possessing
less than about a quarter the Earth's surface water inventory could
rapidly find themselves into a water-trapped configuration, following
their capture into 1:1 spin-orbit resonance. The critical amount of
surface water below which a water-trapped configuration can occur will
probably vary with the magnitude of insolation received by such
planets, although this issue cannot be easily disentangled from the
atmospheric CO$_2$ content and the climate being regulated by
global-scale weathering, which has a potentially strong dependence on
the continental distribution for tidally-locked planets (Edson et
al. 2011).

Clearly, the main argument constructed in this work relies on a number
of assumptions and approximations made in the climate and ice sheet
models, which all deserve further scrutiny. For example, oceanic heat
transport to the nightside, which is not accounted for in either type
of models, could provide an extra heat source that may act to limit
the ice thickness relative to the results discussed
above.\footnote{Under particular conditions, tidal dissipation could
  contribute an additional heat flux acting to limit the nightside ice
  layer thickness (e.g., Barnes et al. 2013; Leconte et al. 2013).} In
the limit where this transport is efficient, significant ice
accumulation may only occur on the fraction of the planetary nightside
that is covered by continents,\footnote{I note that the literature on
  Earth's True Polar Wander suggests that continental distribution and
  planetary moments of inertia are not strongly coupled attributes
  (e.g., Kirschvink et al. 1997, Raub et. al. 2007). This also
  suggests a lack of correlation between the continental distribution
  and the day/night hemispheres that end up being locked into 1:1
  spin-orbit resonance by tides on a close-in planet.} thus reducing
the size of the ice trap. This is an example of how the detailed
topography and ocean basin configuration on a water-trapped world can
have a strong impact on its climate.

Another example of the important role played by the continental
distribution and ocean basin configuration concerns the amount of
residual liquid water left on the dayside of a water-trapped world. In
the climate models discussed in \S\ref{sec:climate}, the ability to
precipitate snow at the rates listed in Table~\ref{tab:one} largely
relies on evaporation over oceans and moisture transport to the
nightside by the atmosphere. In a water-trapped configuration, as the
dayside oceans gradually dry up, their evaporative surface will shrink
in a way that depends on the detailed ocean basin configuration. The
atmospheric moisture transport to the nightside may also be affected
when a significant lowering of the dayside sea level occurs. If this
were to result in substantially decreased nightside precipitation
rates, the nightside ice sheet would tend to accumulate less ice than
evaluated in \S\ref{sec:ice} (see, e.g., scaling with $a$ in
Eq.~(\ref{eq:iceZ})). Eventually, the ice layer depth would not be
limited by melting at the base but simply by mass balance at very
small accumulation rates. This regime, which would require snow
precipitation rates orders of magnitudes smaller than the values
listed in Table~\ref{tab:one}, may happen on some water-trapped worlds
but not others, when dayside oceans dry up. Regardless, this reasoning
illustrates how a reliable evaluation of the amount of liquid water
left on the dayside of a water-trapped world must rely on a closed
hydrological cycle, i.e. one that accounts for two-way exchanges
between the dayside and the nightside water reservoirs, in a manner
more explicit than the isolated ice sheet models considered in
\S\ref{sec:ice}.

\begin{figure*}[l]
\centering \includegraphics[scale=0.11]{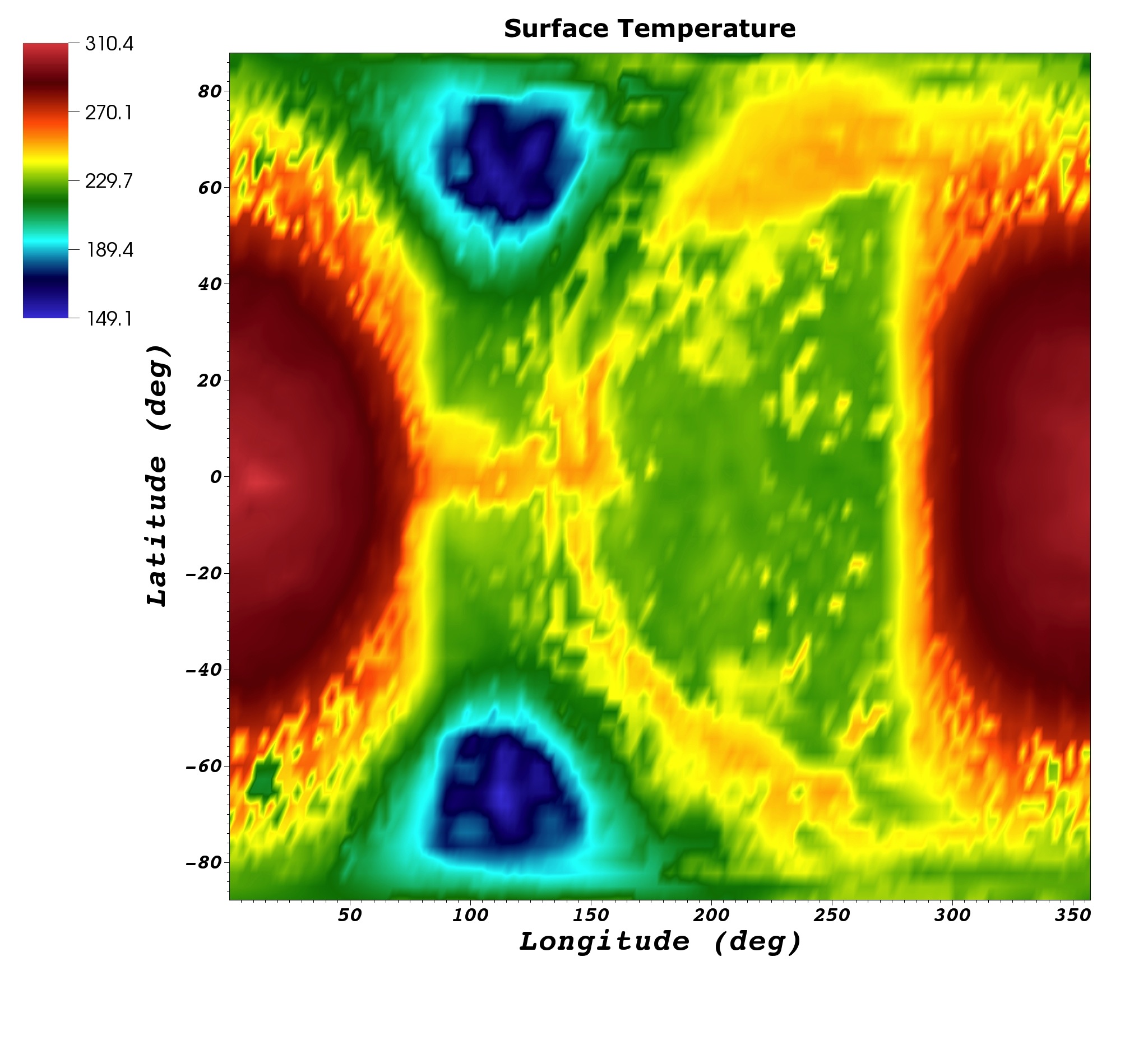}
\centering \includegraphics[scale=0.11]{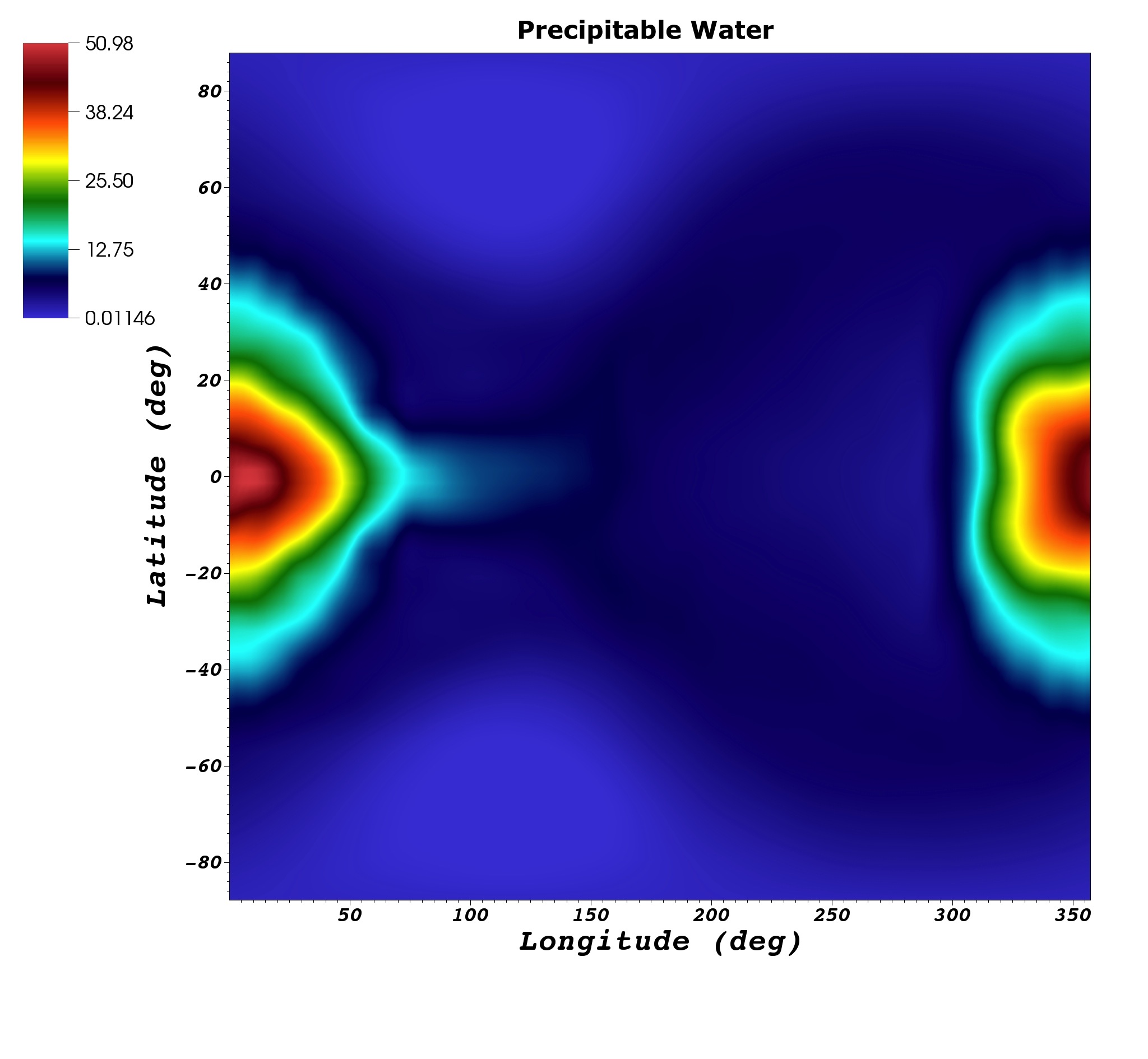}
\centering \includegraphics[scale=0.22]{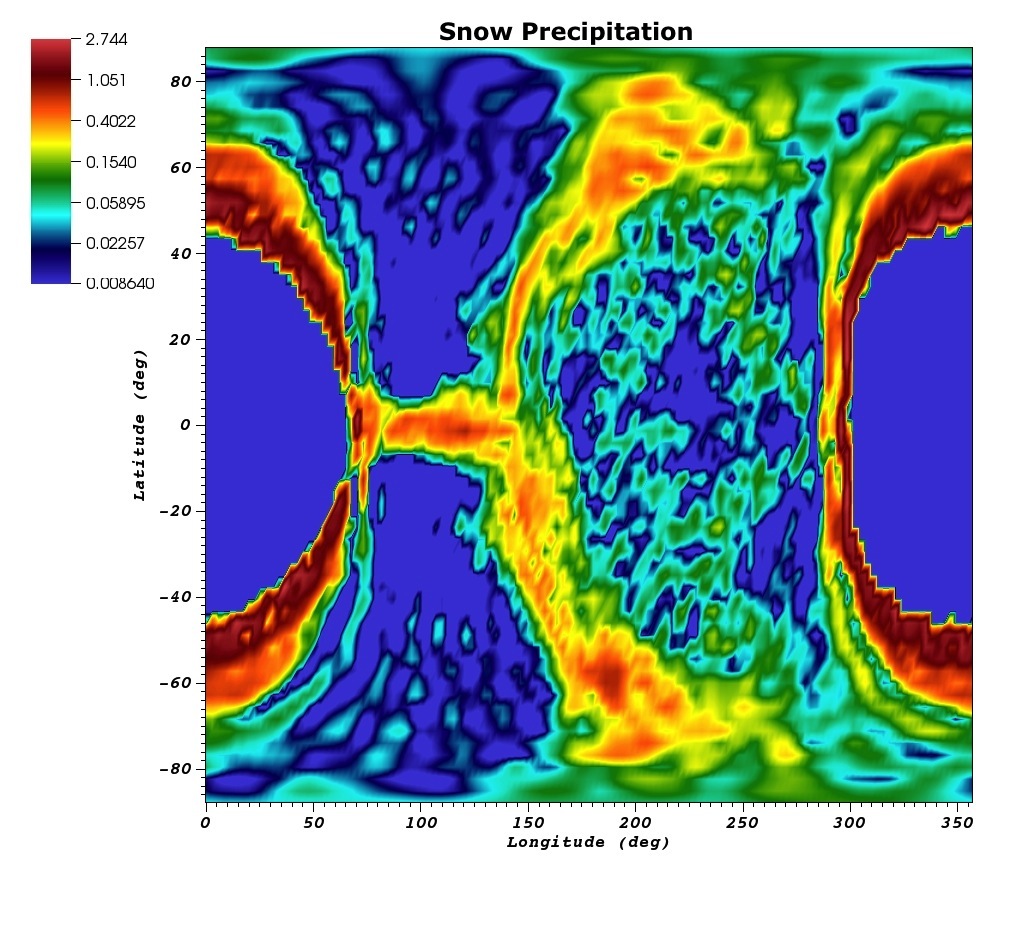}
\centering \includegraphics[scale=0.11]{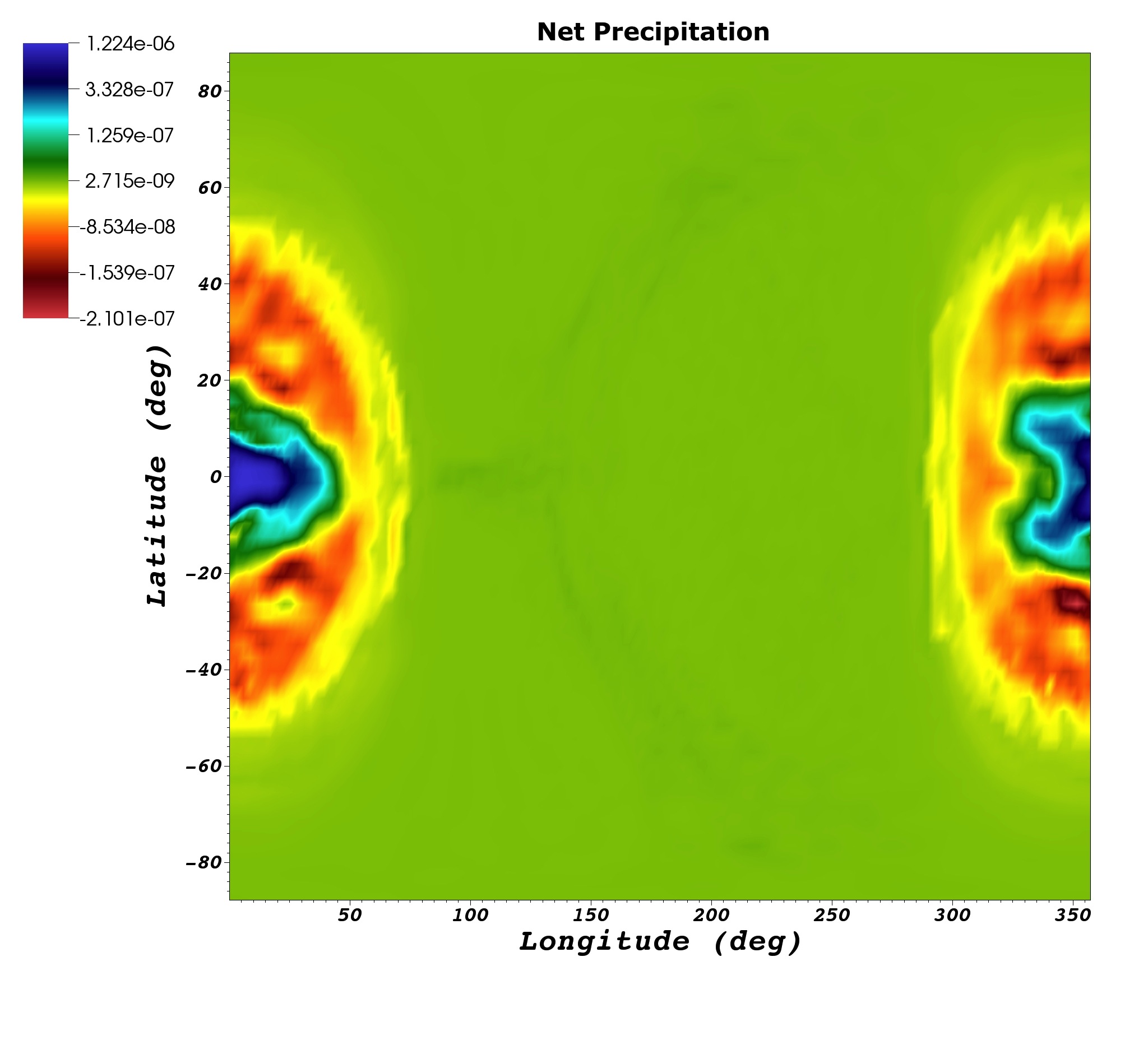}
\caption{End snapshots from the Aquaplanet
  model. Clockwise, from upper left: surface temperature (K),
  precipitable water (kg m$^{-2}$), net precipitation (precipitation
  minus evaporation in m s$^{-1}$, with an arbitrarily skewed scale to
  highlight low values) and snow precipitation (mm day$^{-1}$,
  logarithmic scale with zero reset at $8.64 \times 10^{-3}$). The substellar
  point is located at zero longitude and latitude in each plot.
  Surface water conditions are confined to the dayside and the coldest
  regions on the nightside are offset from the poles. A modest amount
  of moisture is advected to the nightside, resulting in weak snow
  precipitation rates. Net evaporation occurs on the dayside in an
  annulus from $\sim 20$~deg to $\sim 50$~deg from the substellar
  point, with net precipitation over the remainder of the globe.  }
\label{fig:one}
\end{figure*}

The amount of residual liquid water on the dayside of water-trapped
worlds will also depend on the efficiency of rock weathering at
regulating the CO$_2$ content of the atmosphere. Abe et al. (2011)
argue that weathering will continue to operate on land planets but the
efficiency of the process may differ from what it is on Earth,
depending on the availability of surface water for different land
regions.\footnote{For example, desertification could occur in regions
  where evaporation dominates over precipitation, such as the dayside
  annular region away from the substellar point with negative net
  precipitation in Fig.~\ref{fig:one}.} On a planet with dried-up
dayside ocean basins, more rock will be exposed to the atmosphere but
the opportunity for ocean weathering would also largely disappear,
with unclear consequences for the global weathering rate (e.g., Abbot,
Cowan \& Ciesla 2012). Quantitatively, the models Earth.lowCO2 and
Earth.highCO2 listed in Table~\ref{tab:one} exhibit $\sim 150$~m
differences in  their ice equivalent thicknesses, relative to the model
Earth1 with 360ppm of CO$_2$. This makes it clear that even modest
adjustments to the atmospheric CO$_2$ content by weathering can have a
significant impact on the residual dayside surface water inventory of
a water-trapped world.

A related issue is the fate of the large CO$_2$ reservoir stored in
the ocean (about $50$ times that in the atmosphere on Earth). Although
this reservoir may shrink in proportion to the total ocean mass on
volatile deficient planets, if one assumes a fixed CO$_2$ ocean
concentration, it could still amount to a significant and sudden
CO$_2$ increase if released in the atmosphere when dayside oceans
dry-up on a water-trapped world. This excess CO$_2$ would presumably
be weathered on long-enough timescales, so that the long-term climate
would remain determined by detailed balance between the hydrological
cycle and the carbon-silicate cycle.

\begin{figure*}[l]
\centering \includegraphics[scale=0.15]{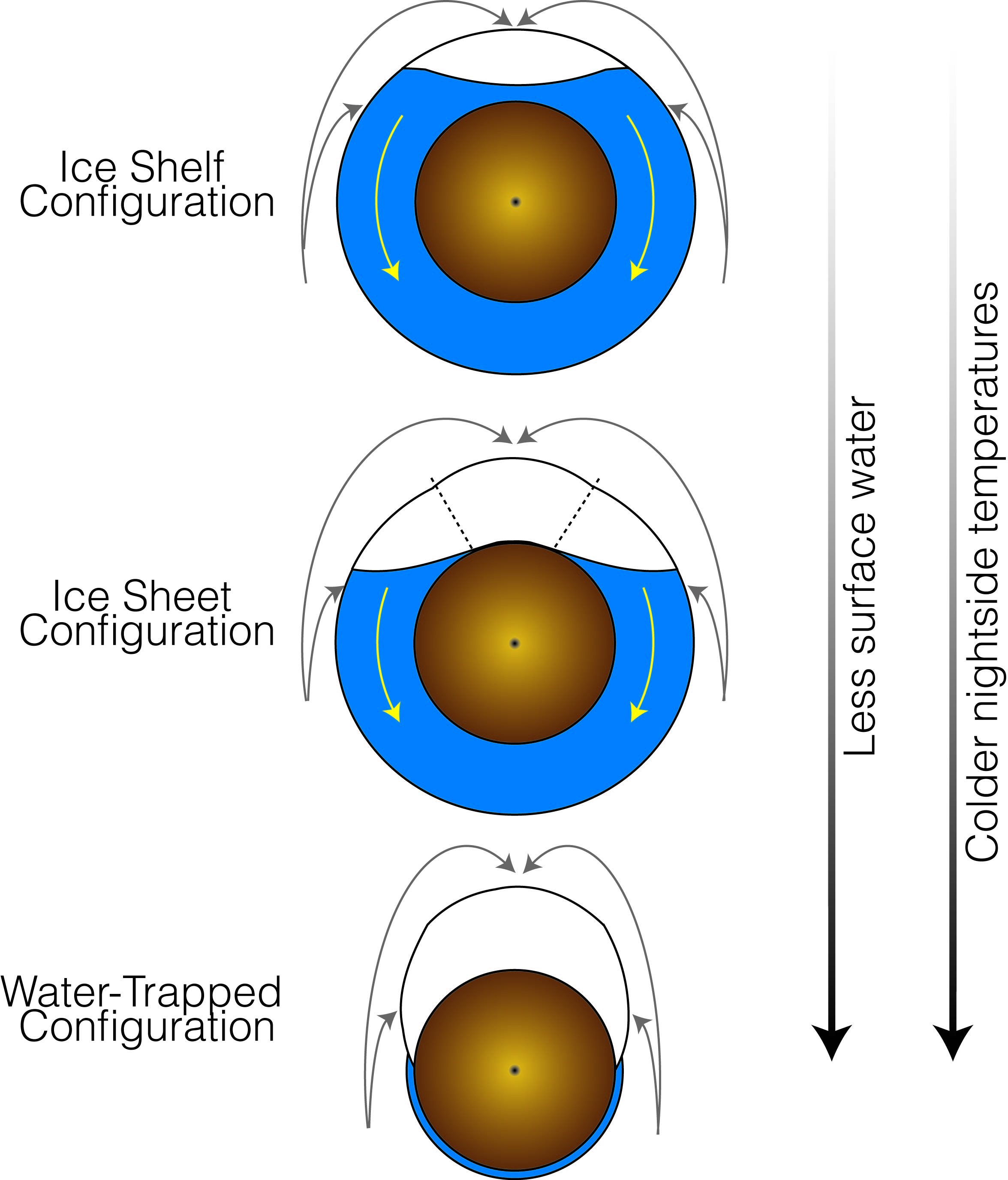}
\caption{Schematic plot of three idealized configurations envisioned
  for the surface water inventory on a tidally-locked terrestrial
  planet. A view from the pole is shown, with the nightside up and the
  dayside down. Exaggerated thicknesses are adopted for the ocean
  (blue) and ice (white) layers. In the ice shelf configuration, the
  ice layer builds up on the nightside but melting occurs at its base
  well before the ground is met.  In the ice sheet configuration, the
  ice layer is partly grounded over the coldest regions on the
  nightside (between the two radial dotted lines) and it connects to
  an ice shelf further away. In the water-trapped configuration, most
  of the surface water is trapped in the nightside ice sheet, with
  comparatively little water on the dayside.  In all cases, the
  hydrological cycle is closed by advection of atmospheric moisture
  from the dayside to the nightside, where it is deposited as snow
  (gray arrows), and a return flow of water to the dayside in the
  ocean (yellow arrows).  Planets with less surface water and/or
  colder nightside temperatures, are more likely to be in the
  water-trapped configuration. }
\label{fig:two}
\end{figure*}

\begin{figure*}[l]
\centering \includegraphics[scale=0.8]{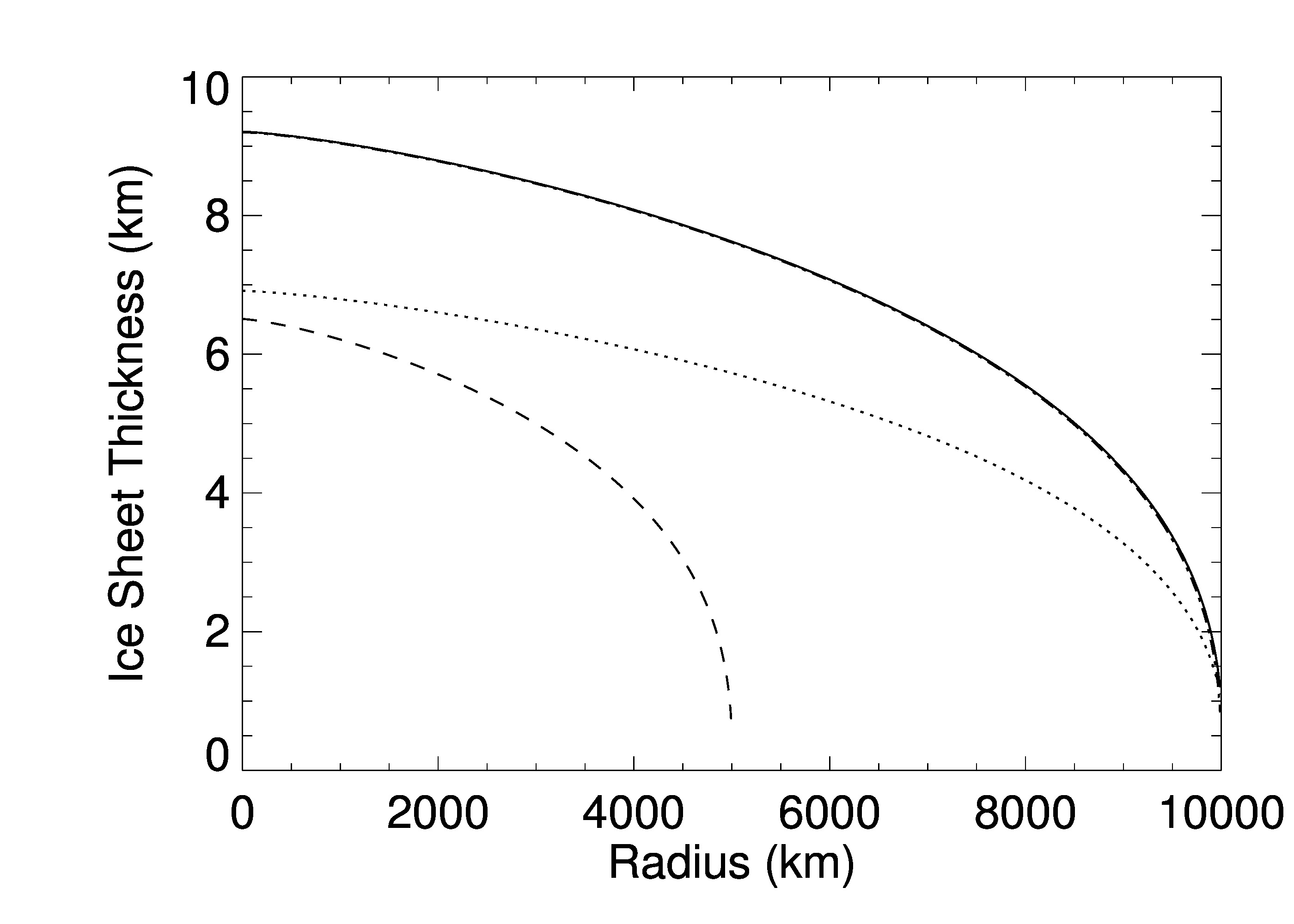}
\caption{Axisymmetric solutions for a thin isothermal flowing ice
  sheet. The solid line is the solution for an ice sheet of $10000$~km
  in radius, a uniform ice accumulation rate of $0.1$~mm day$^{-1}$
  and an outer boundary thickness value of $1$~km.  The dash-dotted
  line, which can hardly be distinguished from the solid line,
  corresponds to a similar solution except for an outer boundary
  thickness value of $500$~m. The dotted line corresponds to a reduced
  ice accumulation rate of $0.01$ mm day$^{-1}$. The dashed line is
  the solution for an ice sheet of $5000$~km in radius (with
  accumulation rate of $0.1$~mm day$^{-1}$ and outer boundary
  thickness value of $500$~m). These steady-state ice profiles carry
  an outward flux of ice at each radius that is equal to the
  integrated ice accumulation rate within that radius.}
\label{fig:three}
\end{figure*}

\begin{figure*}[l]
\centering \includegraphics[scale=0.2]{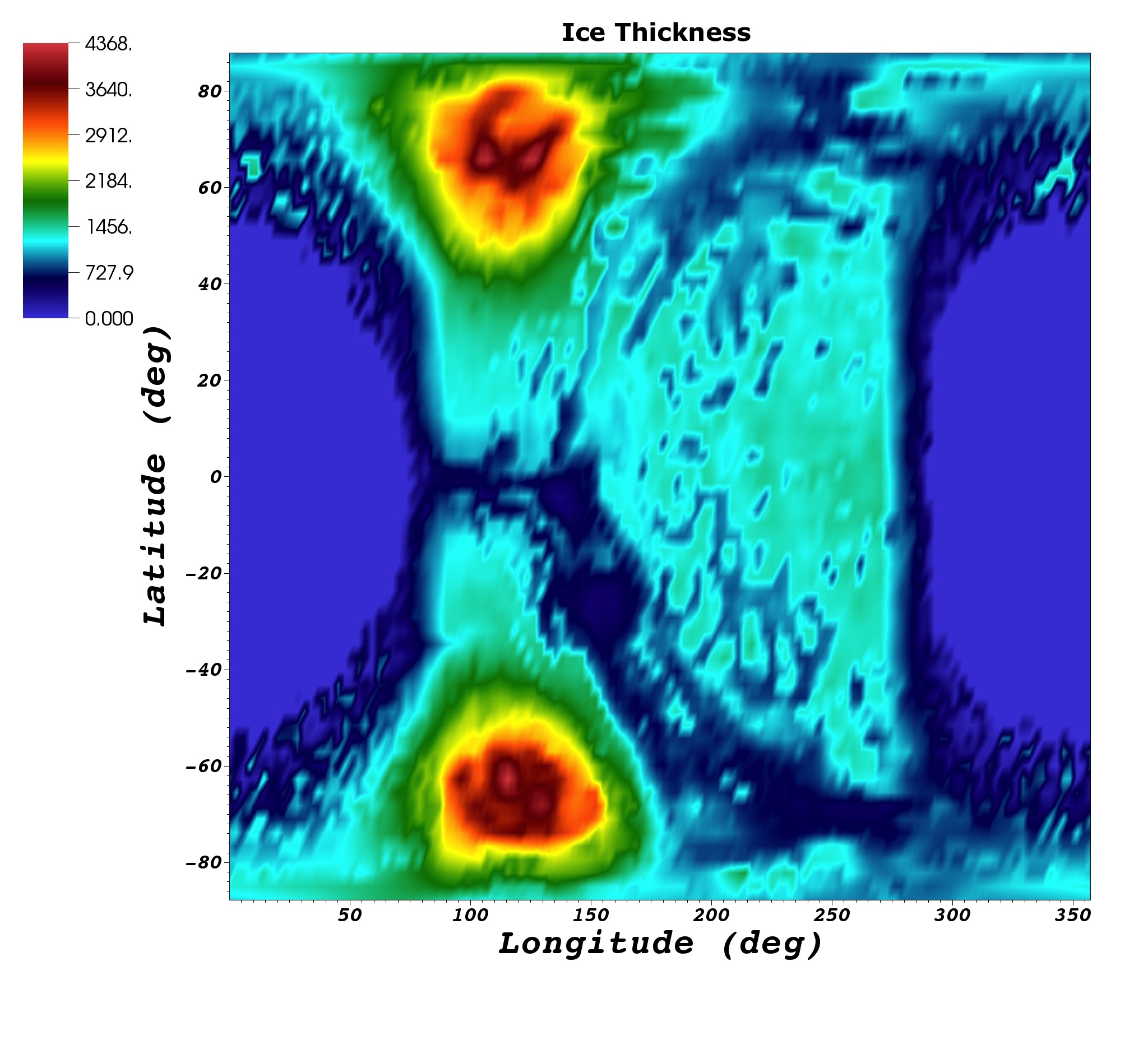}
\caption{A representative map of the ice thickness distribution (in
  meters) permitted by the thermodynamic melting constraint in the
  Aquaplanet model, for a geothermal flux equal to that of Earth. Most
  of the nightside ice has a maximal depth $\sim 1$--$1.5$~km, with a
  colder region of a few thousand km in extent permitting a thicker
  ice layer (up to $4.3$~km). The equivalent depth of this ice
  distribution, averaged over the entire globe, is $560$~m.}
\label{fig:four}
\end{figure*}

The trapping of the majority of the surface water inventory in the
form of nightside ice does not necessarily constitute a challenge to
the habitability of water-trapped worlds. Abe et al. (2011) found more
extended habitability limits for land planets than for aquaplanets and
they illustrated how land planets are better at holding on to their
water, which is lost through atmospheric escape, even if they start
with less water than aquaplanets (see also Leconte et al. 2013; Zsom,
Seager \& de Wit 2013).  Water-trapped worlds with dry daysides and
frozen nightsides may thus offer similar advantages as land planets
for habitability, in comparison with worlds where more abundant water
is free to flow around the globe (first configuration shown in
Fig.~\ref{fig:two}).  A water-trapped configuration may, however, lead
to more strongly localized habitable regions, depending on the
availability and detailed distribution of surface water on the
dayside. It would thus be interesting to determine to what extent
water-trapped and water-free configurations can be discriminated from
one another observationally. This may be challenging if the surface
conditions are not directly accessible (e.g., Benneke \& Seager 2012)
and if the stratospheric water content probed by transmission
spectroscopy is not too different between these two configurations
(e.g., Fig.~12 of Joshi 2003).

\acknowledgments

The author is grateful to D. Valencia for many insights on the
geophysics of this problem, to C. Lithgow-Bertelloni for useful
exchanges on the True Polar Wander literature and to the referee
D. Abbot for many constructive comments, including the suggestion that
thicker ice may be allowed with sub-glacial water flows. This work was
supported in part by NASA grant PATM NNX11AD65G.

\newpage

\begin{table}[h!]
{\tiny
\begin{tabular}{ | c | c | c | c | }
 \hline                       
Model & Key Parameters & Nightside Snow  & Equivalent  Ice  \\
      & & Precipitation (mm/day) & Thickness (m) \\
\hline
Earth1 & $P_{\rm rot}=240$~h, Stellar longitude = $0$~deg & 0.12 & 470 \\
Earth2 & $P_{\rm rot}=240$~h, Stellar longitude = $+90$~deg & 0.11 & 440\\
Earth3 & $P_{\rm rot}=240$~h, Stellar longitude = $+180$~deg & 0.26 & 400\\
Earth.lowP & Earth1 except $P_{\rm surf}=0.3$~bar & 0.27 & 770\\
Earth.lowCO2  & Earth1 except ${\rm CO2}=36$~ppm & 0.10 & 600 \\
Earth.highCO2 & Earth1 except ${\rm CO2}=3600$~ppm & 0.16& 320 \\
Earth.fast & Earth1 except $P_{\rm rot}=120$~h& 0.29 & 340\\
Earth.slow & Earth1 except $P_{\rm rot}=480$~h& 0.009 & 490\\
SuperEarth & Earth1 except $g=29.63$~m~s$^{-2}$, $R_{\rm p} = 11000$~km& 0.07 & 400 \\
Aquaplanet & Earth1 without continents & 0.06 & 560 \\ 
\hline
\end{tabular}
\caption{Key parameters of climate simulations and results.}
\label{tab:one}
}
\end{table}

\newpage

\end{document}